\documentclass[%
 reprint,
 amsmath,amssymb,
 aps,
 prl,
 floatfix,
]{revtex4-1}

\usepackage{graphicx}
\usepackage{dcolumn}
\usepackage{bm}

\usepackage[export]{adjustbox}

\usepackage{url}
\usepackage{subfig}
\usepackage{xcolor}
\usepackage{graphicx}
\usepackage{chemfig}
\newcommand{\materialsN}{11,512} %
\newcommand{\change}[1]{\textcolor{black}{#1}}

\usepackage[normalem]{ulem}

\begin{document}

\preprint{APS/123-QED}

\title{\change{Towards }Novel Organic High-$T_\mathrm{c}$ Superconductors: Data Mining using Density of States Similarity Search}

\author{R.~Matthias~Geilhufe$^1$}
\email{geilhufe@kth.se}
\affiliation{$^1$Nordita, KTH Royal Institute of Technology and Stockholm University, Roslagstullsbacken 23, SE-106 91 Stockholm, Sweden\\
$^2$Department of Computer Science, KTH Royal Institute of Technology, SE-10044 Stockholm, Sweden\\
$^3$Institute for Materials Science, Los Alamos National Laboratory, Los Alamos, NM 87545, USA
}
\author{Stanislav~S.~Borysov$^{1}$}
\email{borysov@kth.se}
\affiliation{$^1$Nordita, KTH Royal Institute of Technology and Stockholm University, Roslagstullsbacken 23, SE-106 91 Stockholm, Sweden\\
$^2$Department of Computer Science, KTH Royal Institute of Technology, SE-10044 Stockholm, Sweden\\
$^3$Institute for Materials Science, Los Alamos National Laboratory, Los Alamos, NM 87545, USA
}
\author{Dmytro~Kalpakchi$^{1,2}$}
\affiliation{$^1$Nordita, KTH Royal Institute of Technology and Stockholm University, Roslagstullsbacken 23, SE-106 91 Stockholm, Sweden\\
$^2$Department of Computer Science, KTH Royal Institute of Technology, SE-10044 Stockholm, Sweden\\
$^3$Institute for Materials Science, Los Alamos National Laboratory, Los Alamos, NM 87545, USA
}
\author{Alexander~V.~Balatsky$^{1,3}$}
\email{balatsky@hotmail.com}
\affiliation{$^1$Nordita, KTH Royal Institute of Technology and Stockholm University, Roslagstullsbacken 23, SE-106 91 Stockholm, Sweden\\
$^2$Department of Computer Science, KTH Royal Institute of Technology, SE-10044 Stockholm, Sweden\\
$^3$Institute for Materials Science, Los Alamos National Laboratory, Los Alamos, NM 87545, USA
}

\date{\today}

\begin{abstract}
\change{Identifying novel functional materials with desired key properties is an important part of bridging the gap between fundamental research and technological advancement. In this context, high-throughput calculations combined with data-mining techniques highly accelerated this process in different areas of research during the past years. The strength of a data-driven approach for materials prediction lies in narrowing down the search space of thousands of materials to a subset of prospective candidates. Recently, the open-access organic materials database OMDB was released providing electronic structure data for thousands of previously synthesized three-dimensional organic crystals. Based on the OMDB, we report about the implementation of a novel density of states similarity search tool which is capable of retrieving materials with similar density of states to a reference material. The tool is based on the approximate nearest neighbor algorithm as implemented in the ANNOY library and can be applied via the OMDB web interface. The approach presented here is wide-ranging and can be applied to various problems where the density of states is responsible for certain key properties of a material. As the first application, we report about materials exhibiting electronic structure similarities to the aromatic hydrocarbon p-terphenyl which was recently discussed as a potential organic high-temperature superconductor exhibiting a transition temperature in the order of 120~K under strong potassium doping. Although the mechanism driving the remarkable transition temperature remains under debate, we argue that the density of states, reflecting the electronic structure of a material, might serve as a crucial ingredient for the observed high-$T_\mathrm{c}$. To provide candidates which might exhibit comparable properties, we present 15 purely organic materials with similar features to p-terphenyl within the electronic structure}, which also tend to have structural similarities with p-terphenyl such as space group symmetries, chemical composition and molecular structure. \change{The experimental verification of these candidates might lead to a better understanding of the underlying mechanism in case similar superconducting properties are revealed.}
\end{abstract}

\maketitle

\section{Introduction}\label{sec:intro}
Since the first discovery of superconductivity within organic materials in the late 1970s \cite{jerome1980superconductivity}, the study of organic conductors and their transition to a superconducting state has become an active branch in condensed matter physics. In comparison to inorganic materials, most of the three-dimensional organic crystals exhibit large band gaps \cite{borysov2016}, which makes the identification of organic metals a highly non-trivial task. Among the first reported and widely discussed organic superconductors (e.g., TMTSF- or BEDT-TTF-based materials) are quasi one- or two-dimensional narrow band gap semiconductors which show metallic conductivity under high pressure \cite{kino2006first,kondo2009crystal}. The commonly unconventional pairing mechanisms within organic materials lead to symmetries of the superconducting gap beyond the BCS $s$-wave, such as a $d$-wave gap in $\kappa$-(BEDT−TTF)$_2$Cu[N(CN)$_2$]Br \cite{PhysRevB.61.7033} and $\kappa$-(BEDT-TTF)$_2$Cu(NCS)$_2$ \cite{nam1999angle}, \change{and even odd-frequency superconductivity measured in (TMTSF)$_{2}$ClO$_{4}$ \cite{PhysRevLett.110.107005} and theoretically predicted in $\kappa$-(BEDT-TTF)$_2$X \cite{vojta1999indications}}. For both the pressurized organic semiconductors as well as the pure organic metals, the superconducting transition temperature is usually \change{in the order of 10~K }\cite{lang2004organic}. 

In contrast to the small transition temperatures observed so far, it has been recently reported that potassium-doped p-terphenyl exhibits a superconducting transition temperature of 123~K which is comparable to transition temperatures of high-temperature superconducting cuprates \cite{wang2017superconductivity}. This remarkable finding might open a path towards high-temperature organic superconductivity. To extend the discussion towards other materials candidates, we report about the results of a data mining exercise for other organic crystals with similar electronic features as p-terphenyl. \change{To date, the mechanism driving the high transition temperature remains unclear. }

The data mining is performed on electronic structure data of \materialsN{} organic crystals stored within the open-access Organic Materials Database (OMDB) \cite{borysov2016}, freely accessible at \url{http://omdb.diracmaterials.org}. It contains results of band structure and density of states (DOS) calculations for previously synthesized three-dimensional organic crystals, where the crystal structures are mainly taken from the Crystallographic Open Database (COD) \cite{downs2003american,gravzulis2009crystallography,gravzulis2012crystallography,gravzulis2015computing,merkys2016cod}. The web interface of the OMDB contains advanced search functionality such as pattern matching for specific features within electronic band structures \cite{band_tool} and the newly developed DOS similarity search tool presented in this paper. 

Recently, a data mining approach was successfully applied, for example, for the search of stable nitride perovskites~\cite{sarmiento2015prediction}, thermoelectric materials \cite{PhysRevX.1.021012} and Dirac materials \cite{klintenberg2014computational,geilhufe_point,geilhufe_line,yan2017data}. In a similar manner as reported in the current paper, potential inorganic high-temperature superconductors were previously reported by Klintenberg and Eriksson \cite{klintenberg2013possible} after comparing electronic structure features of the materials stored within the Electronic Structure Project \cite{ortiz2009data} to certain prototype materials. 

Even though important contributions towards \emph{ab initio} description of superconductivity were achieved \cite{luders2006density,lueders2005,marques2005}, a general theory for high-$T_\mathrm{c}$ superconductors is still under debate. However, since the normal state electronic properties together with the structure of the material will significantly influence the occurrence of a superconducting phase, a similarity search for distinct features with respect to a prototype material represents a valid approach to guide the search for other potential high-temperature superconductors. \change{The results presented here are based on a high-throughput approach, intended to provide guidance for a experimental verification of prospective candidates. A positive experimental verification of high-temperature superconductivity within the 15 materials suggested in the following will motivate a specific microscopic description of the superconducting state.}

\change{
The paper is structured as follows. At the beginning, we give details on the implementation of the novel DOS similarity search tool within the OMDB. Afterwards, we illustrate the electronic structure of the prototype material p-terphenyl which was calculated in the framework of the density functional theory (DFT). Finally, we present results of the similarity search within the OMDB and list the 15 materials which are most similar to p-terphenyl.  
}

\section{DOS similarity search tool}\label{sec:search_tool}
\change{We developed a DOS similarity search tool which is freely available online at \url{https://omdb.diracmaterials.org/search/dos}. It allows to retrieve materials from the OMDB with a DOS similar to the one of a provided reference material. The DOS search tool implementation follows the same line as the previously developed pattern search tool for electronic band structures within the OMDB \cite{band_tool}. The implementation uses the approximate nearest neighbor search algorithm and is based on the ANNOY library \cite{annoy}. Subsequently, the implementation is summarized.}

\change{\textit{The interface.} At the current stage, the reference DOS uploaded by the user has to be either in the JSON format or in the DOSCAR format, which is the standard output of the VASP code \cite{vasp1,vasp2,vasp3}. An extension to other standard formats used within major \emph{ab initio} codes will be provided in the nearest future. Within the reference DOS, the user specifies the part of interest, which is used as the search query. The specified part also defines the size of the moving window used for the search. Furthermore, an energy search range needs to be specified, which can either comprise the entire DOS available for all stored materials or it can be fixed to a specific value, as the minimum energy of the lowest conduction band as a lower bound or the maximum energy of the highest valence band as an upper bound of the search window. 
}
\begin{figure}[t]
\includegraphics[width=9cm]{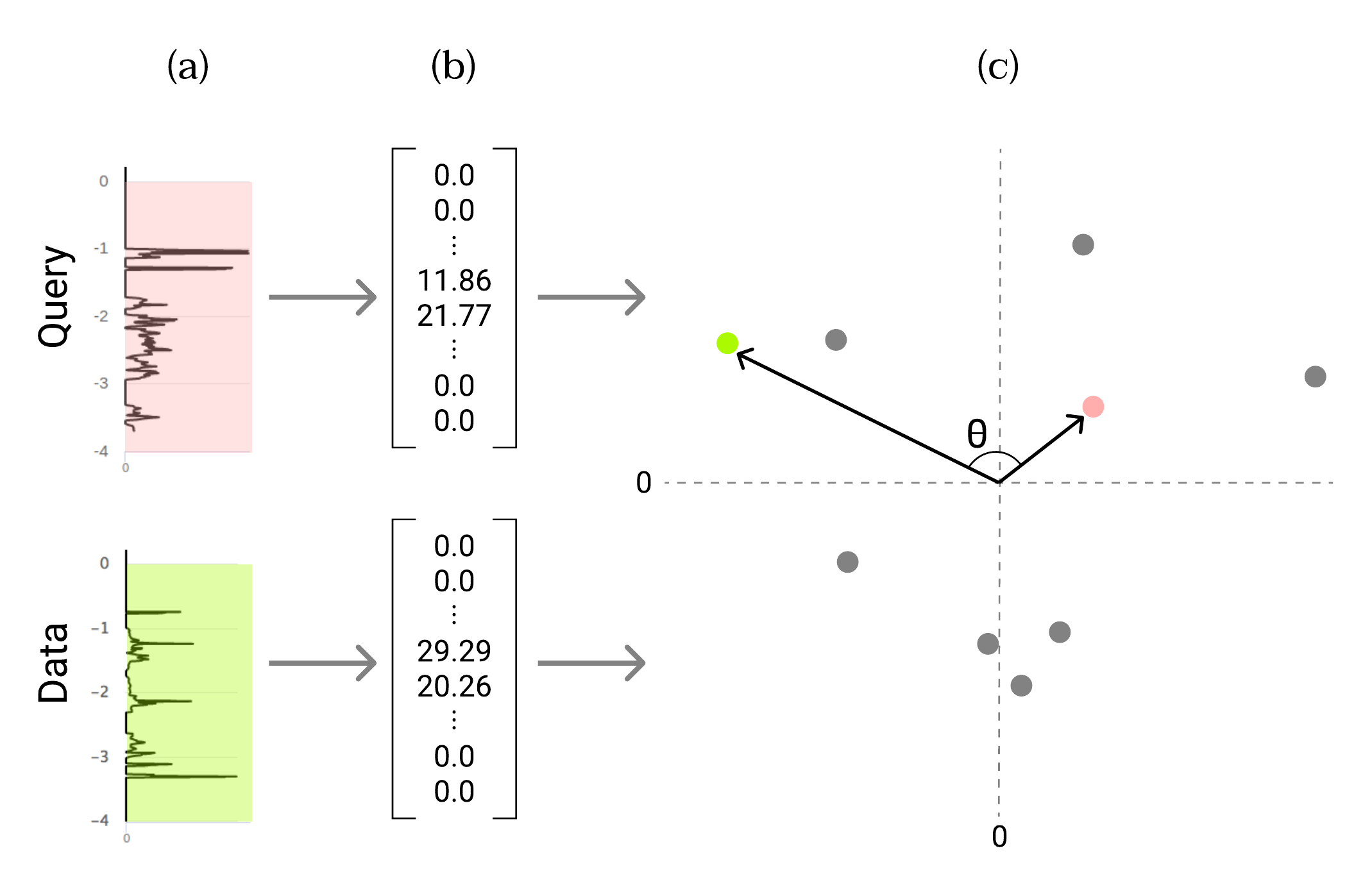}
\caption{\change{Calculation of similarity between two DOS. (a) the reference DOS to be used as a query and a part of DOS  of the same length in the energy space for a material from the database are represented as high-dimensional vectors (b). Similarity between these two DOS is based on the angle $\theta$ between the corresponding vectors (c), where $\theta=0$ corresponds to a perfect match (two DOS are equivalent up to a scaling constant). Here, a principal component projection from the high-dimensional to a two-dimensional space is used for visualization purposes.
}\label{fig:dos_search}}
\end{figure}

\change{\textit{Search algorithm.}} Given the query DOS of length $w$, we employ a moving window approach, where DOS contained within the OMDB are scanned using a moving window of the same size $w$ and stride $s$, which defines the number of data points the window jumps at each scanning step. Since $s$ should be small with respect to $w$ (which is of the order of eV), we set $s=0.02$~eV. In the case the user specifies a fixed search bound, e.g., corresponding to the maximum energy of the highest valence band, only one search window for each material is used. 

Both the query DOS and the data points contained within the moving window \change{(Fig.~\ref{fig:dos_search}a)} are represented as \change{high-dimensional} vectors of size $d$ \change{(Fig.~\ref{fig:dos_search}b)} where a linear interpolation is applied to achieve the required number of equidistant points. As $d$ should be large enough for a vector to represent DOS patterns, we define $d=w/s+1$ (which corresponds for example to 51 points for $w=1$~eV). The similarity between the normalized query and window vectors is measured using the cosine distance $\sqrt{2-2\cos{\theta}}$, where $\theta$ is the angle between the vectors \change{(Fig.~\ref{fig:dos_search}c)}. As $\theta$ ranges from $0$ (two vectors are \change{equivalent up to a scaling factor}) to $\pi$ (two vectors \change{have opposite directions}), the cosine distance ranges from 0.0 to 2.0, respectively.


\change{Due to high computational costs, a linear brute force solution, i.e., a sequential comparison of all data vectors to a search query, is not suitable for online implementations, where results are expected within a short amount of time. Therefore, we use the approximate nearest neighbor search algorithm as implemented within the ANNOY library. During an indexing step, ANNOY creates a forest of binary search trees by applying the random projection method, partitioning the $d$-dimensional search space.} During the search step, the trees allow to find the closest vectors to a query vector within logarithmic time. Since this performance improvement comes in the expense of increased memory usage, only moving windows of 1~eV and 2~eV are indexed at the current stage and available for search. More details about the approximate nearest neighbor search algorithm and the ANNOY library can be found in Ref.~\cite{band_tool}.

\section{Reference material p-Terphenyl}
\subsection{Crystal structure and electronic structure}
The organic compound p-terphenyl or 1,4-Diphenylbenzene crystallizes within a monoclinic crystal structure obeying the space group $P2_1/a$ (14). The molecule itself has the chemical sum formula C$_{18}$H$_{14}$, exhibiting the following skeletal formula: 
\begin{center}
\chemfig[scale=0.75][scale=0.75]{[:-30]*6((-*6(=-=-=-))=-=(-*6(=-=-=-))-=-)} 
\end{center}
According to Ref.~\cite{Rietveld:a07448}, the lattice parameters are given by $a=8.10$ \AA, $b=5.61$ \AA, $c=13.61$ \AA~and $\beta=92.02^\circ$. 

\begin{figure}[t]
\subfloat[band structure along various high symmetry paths within the Brillouin zone\label{fig:bands}]{\includegraphics[width=6.2cm]{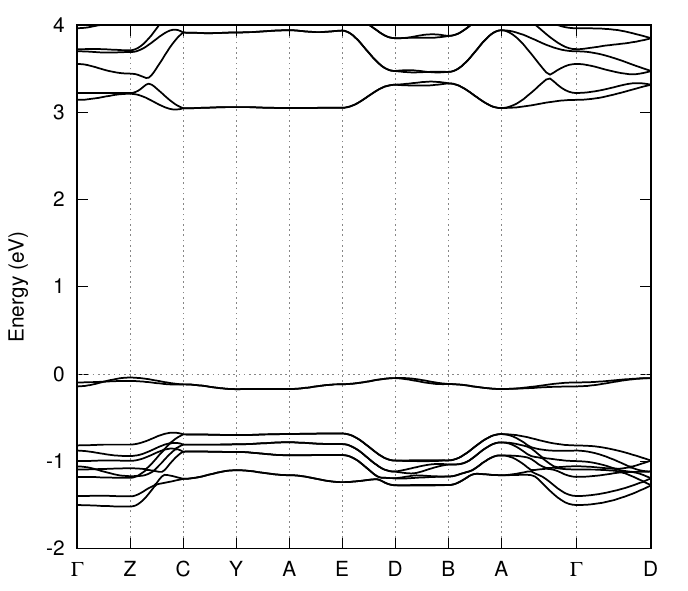}\raisebox{0.34cm}{\includegraphics[width=3.6cm]{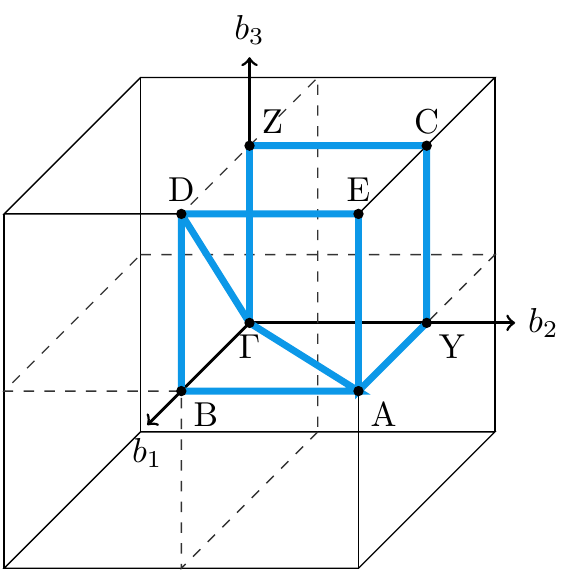}}}\\
\subfloat[density of states including the two energy windows W1 and W2 used for the similarity search\label{fig:dos}]{\includegraphics[width=5cm]{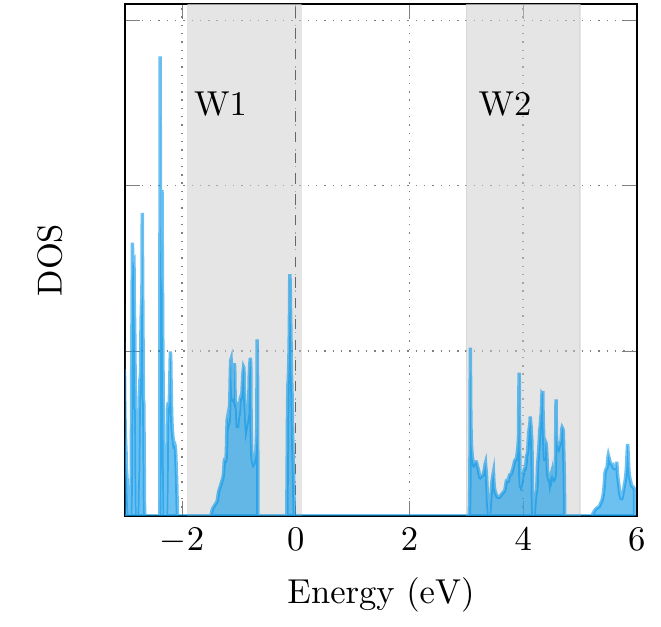}}
\caption{Electronic structure of p-terphenyl.\label{fig:es}}
\end{figure}

The electronic structure of p-terphenyl for a non-optimized unit cell with ionic positions according to Ref.~\cite{Rietveld:a07448} is illustrated in Fig.~\ref{fig:es}. The crystal itself is a large band gap insulator with a calculated band gap of 3~eV. The band structure (Fig.~\ref{fig:bands}) exhibits a localized and well isolated flat band right below the Fermi level which is also seen in the density of states shown in Fig.~\ref{fig:dos}. Such flat bands are common within the band structure of organics due to the weak hopping of electrons between the molecular orbitals. Both below and above the gap, the DOS shows localized features within an energy range of about 2~eV. After a structural optimization of the ionic positions, the band structure and density of states remain qualitatively similar. The band gap itself slightly decreases to about 2.8~eV, whereas the spectral gap between the localized band below the Fermi level to the second energetically lower branch of bands slightly increases to about 0.8~eV.

\subsection{Computational details}
The electronic structure of the reference material p-terphenyl was calculated in the framework of the density functional theory~\cite{Hohenberg1964,Kohn1965,Jones2015} by applying a pseudopotential projector augmented-wave method~\cite{hamann1979norm,blochl1994projector,pseudo1,pseudo2}, as implemented in the Vienna Ab initio Simulation Package (VASP)~\cite{vasp1,vasp2,vasp3,kresse1999ultrasoft}. During the calculations, the exchange-correlation functional was approximated by the generalized gradient approximation according to Perdew, Burke and Ernzerhof~\cite{perdew1996} (PBE). The structural information was taken from the Cambridge Structural Database (CSD) \cite{Groom:bm5086,Rietveld:a07448} and transferred into POSCAR files using VESTA (Visualization for Electronic and STructural Analysis) \cite{momma2011vesta}. 

The energy cut-off was set to 400~eV. Due to the light atoms involved, the calculations were performed spin-polarized but without spin-orbit coupling. For the integration in $\vec{k}$-space, a $6\times6\times6$ $\Gamma$-centered mesh according to Monkhorst and Pack~\cite{monkhorst1976special} was chosen during the self-consistent cycle. Band structure calculations were performed in two ways. First, the crystal structure was kept fixed as provided by the experiment \cite{Rietveld:a07448}, which is in correspondence to the electronic structure calculations stored within the OMDB \cite{borysov2016}. Second, the ionic positions were optimized by incorporating van der Waals corrections within the framework of the density dependent dispersion correction according to Steinmann and Corminboeuf \cite{Steinmann2011} to cross-check the stability of the electronic structure with respect to slight structural changes. 

\section{Search results and discussion}\label{sec:discussion}
\begin{figure}[t]
\subfloat[
OMDB-ID 12336, COD-ID 2212018, C$_8$H$_8$Br$_2$]{
\includegraphics[width=8.cm]{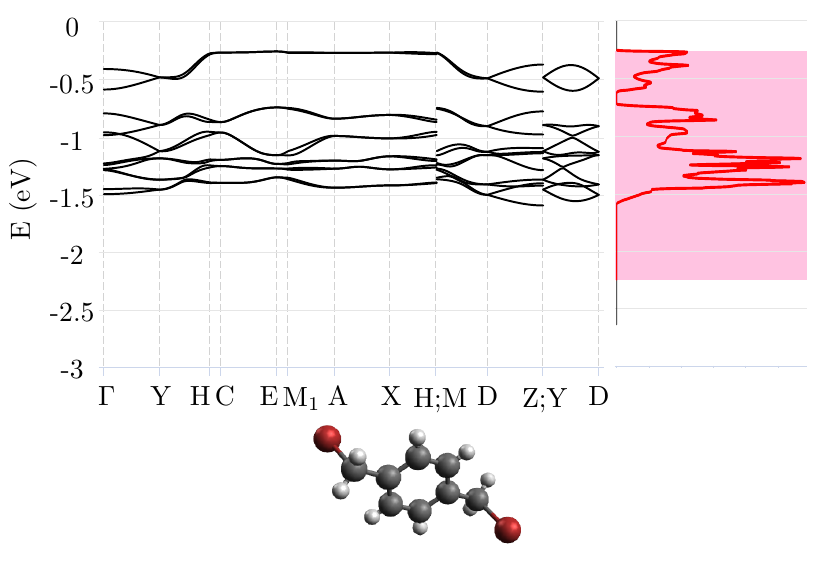}
}\\
\subfloat[OMDB-ID 261, COD-ID 4062949, C$_8$H$_5$O]{
\includegraphics[width=8.cm]{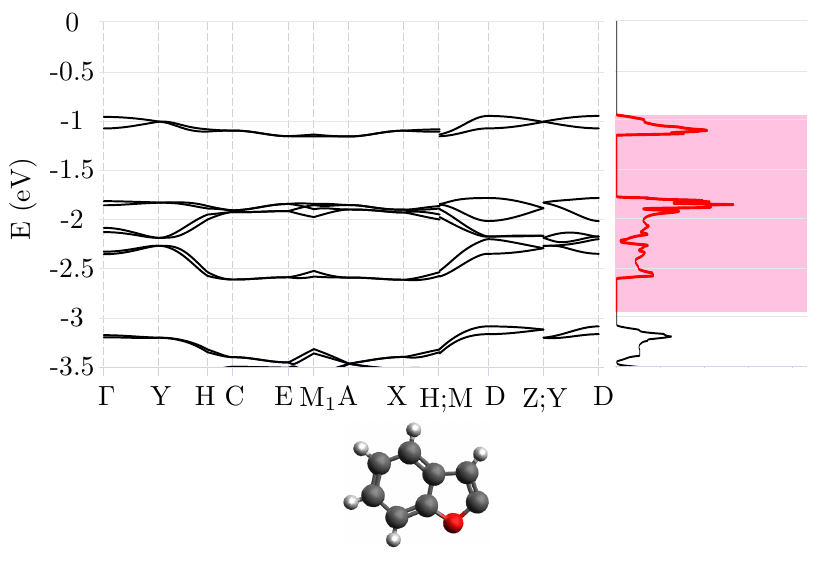}
}\\
\subfloat[
OMDB-ID 12231, COD-ID 2203730, C$_{18}$H$_{18}$O$_2$]{
\includegraphics[width=8.cm]{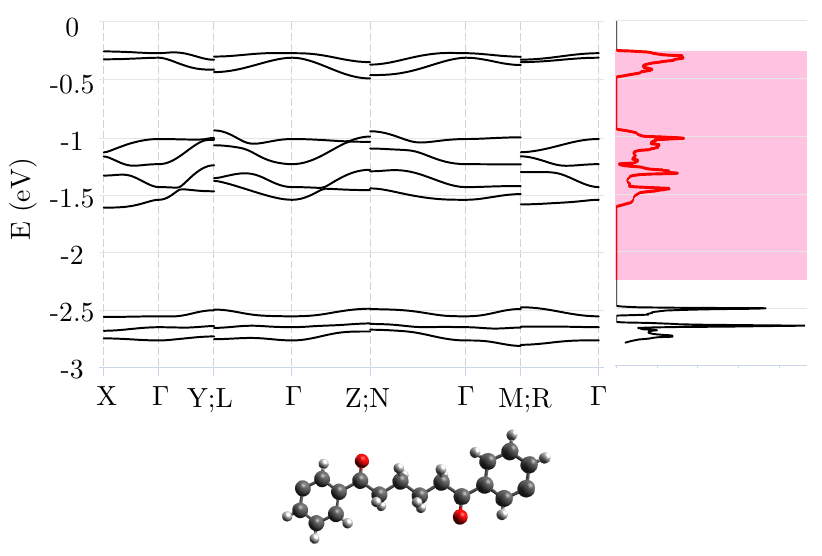}
}
\caption{Electronic structure of the three most promising materials with similarity to the electronic structure of p-terphenyl within the valence bands. 
\label{res:val}}
\end{figure}

\begin{figure}[t]
\subfloat[
OMDB-ID 20042, COD-ID 2222905, C$_{13}$H$_9$FO$_3$]{
\includegraphics[width=8.cm]{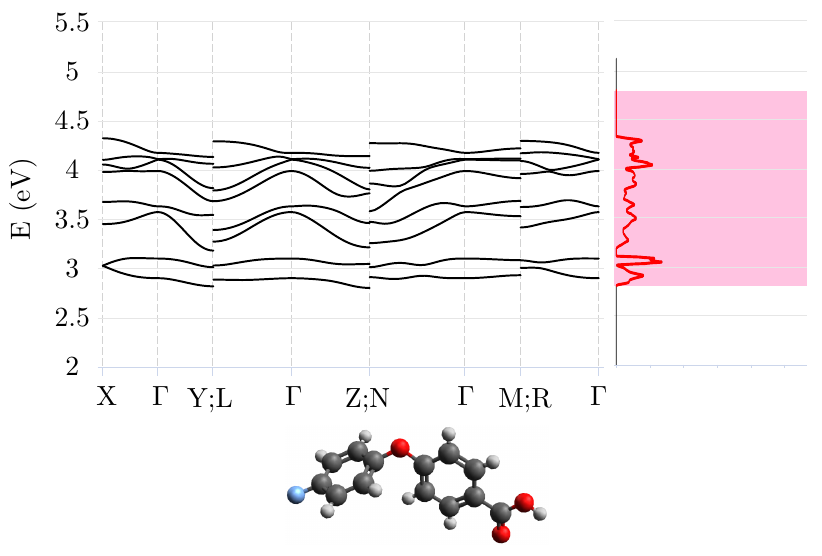}
}\\
\subfloat[OMDB-ID 21466, COD-ID 7153064, C$_{16}$H$_{12}$ClN$_3$O]{
\includegraphics[width=8.cm]{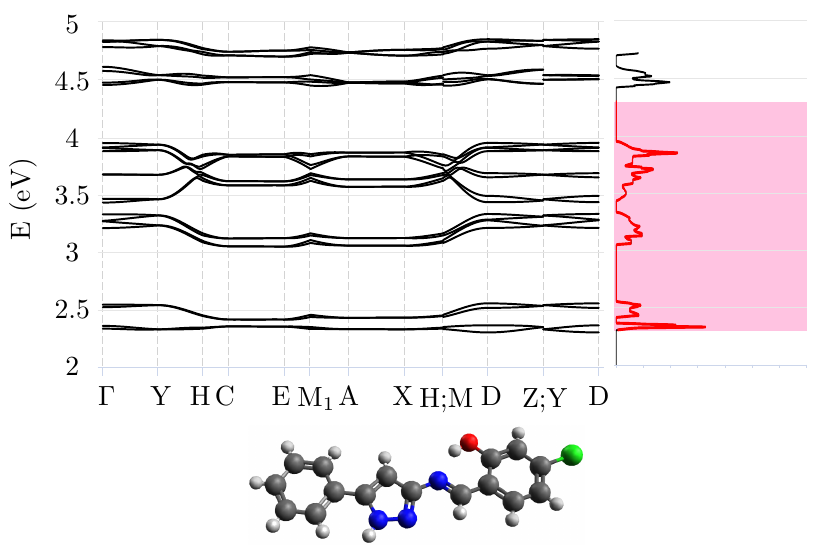}
}\\
\subfloat[OMDB-ID 19325, COD-ID 1542706,C$_{12}$H$_{14}$N$_2$O$_5$]{
\includegraphics[width=8.cm]{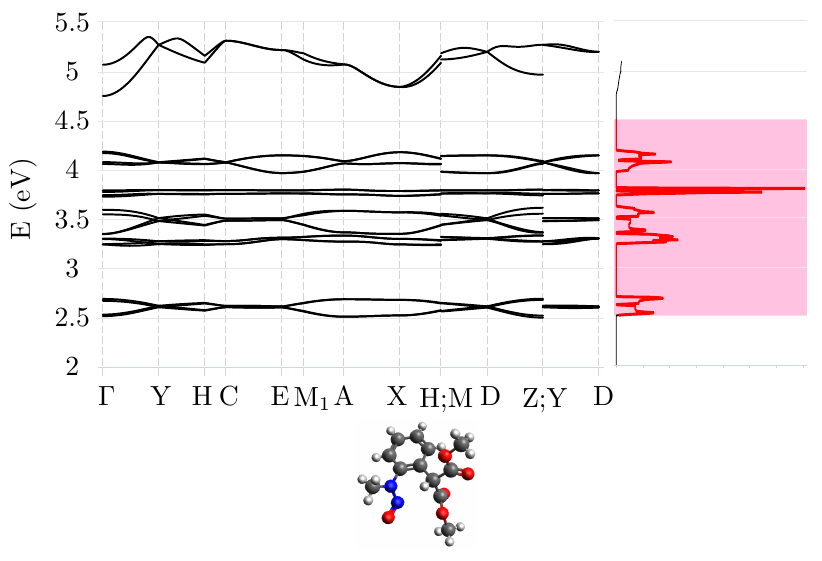}
}
\caption{Electronic structure of the three most promising materials with similarity to the electronic structure of p-terphenyl within the conduction bands. 
\label{res:cond}}
\end{figure}

To identify materials with similar electronic properties as p-terphenyl, the DOS similarity search tool was applied to \materialsN{} materials stored within the OMDB at the moment of writing. For each material, the search was performed separately for the region below the highest occupied state as well as above the lowest unoccupied state. We used the search window size of 2~eV with the maximum energy of the highest valence band / minimum energy of the lowest conduction band as an upper / lower bound of the search window (see also Fig.~\ref{fig:es}). The search within the window below the highest occupied state identifies 16 materials showing a minimal distance of 0.641 and a maximal distance of 0.791 to the reference DOS. The similar search above the lowest unoccupied state gives 5 materials having a minimal distance of 0.641 and a maximal distance of 0.767. The best search results are shown in Table~\ref{Tab:results} and electronic structures of the three top examples are shown in Figs.~\ref{res:val} and \ref{res:cond}, for the search above the highest occupied and lowest unoccupied states, respectively. The best similarity (with a distance of 0.641) was observed for 1,4-Bis(bromomethyl)benzene (OMDB-ID 12336, COD-ID 2212018).  
\begin{table}[t]
\subfloat[Similar materials within valence bands]{
\begin{tabular}{llllll}
\hline\hline
OMDB-ID & COD-ID & Chemical formula & Dist. & Space Group \\
\hline
12336 & 2212018 & C$_8$H$_8$Br$_2$ & 0.641 & $P2_1/c$\\
261 & 4062949 & C$_8$H$_5$O & 0.693 & $P2_1/c$\\
11968 & 2018118 & C$_9$H$_9$Br$_3$ & 0.714 &  $P\overline{1}$\\
12231 & 2203730 & C$_{18}$H$_{18}$O$_2$ & 0.716 &  $P\overline{1}$\\
2731 & 4020760 & C$_{12}$H$_{16}$O$_3$ & 0.722 & $P2_1/c$\\
4909 & 7150359 & C$_5$H$_3$N$_{11}$ & 0.725 & $P2_1/c$\\
2692 & 4028771 & C$_{15}$H$_{20}$N$_2$O & 0.738 & $P2_1$\\
21766 & 8102335 & C$_4$H$_4$O$_4$Pb & 0.751 &  $P2_1/c$\\
17076 & 4505801 & C$_9$H$_9$I$_3$ & 0.762 &  $P\overline{1}$\\
11918 & 2014326 & C$_3$H$_5$NS$_2$ & 0.768 & $P\overline{1}$\\
\hline\hline
\end{tabular}}\\
\subfloat[Similar materials within conduction bands]{
\begin{tabular}{lllll}
\hline\hline
OMDB-ID & COD-ID & Chemical formula & Dist. & Space Group \\
\hline
20042 & 2222905 & C$_{13}$H$_9$FO$_3$ & 0.717 & $P\overline{1}$ \\
21466 & 7153064 & C$_{16}$H$_{12}$ClN$_3$O & 0.756 & $P 2_1/n $ \\
19325 & 1542706	& C$_{12}$H$_{14}$N$_2$O$_5$ & 0.763 & $P 2_1/c $ \\
21484 & 7153556	& C$_{17}$H$_{11}$BrN$_2$O & 0.766 & $P 2_1/c $ \\
17474 & 7118599	& C$_{21}$H$_{19}$P & 0.767 & $P1$ \\
\hline\hline
\end{tabular}}
\caption{Top search results for the DOS similarity search close to the highest occupied and the lowest unoccupied electronic levels.\label{Tab:results}}
\end{table}

In general, a striking correlation between the crystal structure of the identified materials with the reference material can be observed. 8 out of 15 materials exhibit the same space group and one material belongs to the same crystal class. Additionally, 14 of the 15 materials contain aromatic rings and 10 of the 15 materials contain almost planar molecules comparable to p-terphenyl.
%
%
%
%

The mechanisms guiding the high-temperature superconductivity in p-terphenyl are still unclear. The first attempts were reported by Mazziotti \textit{et al.} \cite{mazziotti2017possible} proposing that the driving mechanism is the quantum resonance between different superconducting gaps near a Lifshitz transition.

Considering a BCS-type mechanism to induce superconductivity, the transition temperature $T_\mathrm{c}$ increases with increasing DOS at the Fermi level $N(E_\mathrm{F})$ and coupling constant $g$,
\begin{equation}
 T_c \sim e^{-1/\left|g\right|N(E_\mathrm{F})}.
\end{equation}
As shown before, pristine p-terphenyl is a large band gap insulator. However, according to Ref.~\cite{wang2017superconductivity}, a large superconducting transition temperature of 123~K was observed after doping the material with potassium. By introducing additional electrons into the system, potassium shifts the Fermi level into the conduction band resulting in an $n$-doped system. However, as large peaks due to the flat bands can be observed in both valence and conduction bands, a similar transition temperature might be observed by introducing $p$-doping. In comparison to $n$-doping, $p$-doping is more straightforward to achieve for organic materials \cite{lussem2013doping,Pfeiffer200389}. Similarly to p-terphenyl, localized peaks in the density of states can also be observed in the suggested materials (Table~\ref{Tab:results}). 
Although $p$- and $n$-doping can affect electronic structure, we expect that couplings will depend on its averaged properties. Indeed, the electron-phonon coupling in the Eliashberg approach is integrated over the frequency range. 
If this coupling is responsible for the high $T_\mathrm{c}$ in the original compound, we expect the same mechanisms to be in play for the proposed materials. 
\section{Conclusions}
\change{We described the implementation of the novel density of states similarity search tool based on the approximate nearest neighbor method implemented in the ANNOY package. In connection to the present work, the search tool was released within the web interface of the OMDB, an open-access electronic structure database for previously synthesized organic materials. The approach presented in the current work is wide-ranging and can be applied to various problems where the density of states is (or might be) connected to the properties of interest.}

To propose a guidance in narrowing down the search space for novel organic high-temperature superconductors, \change{we applied the novel search tool to query for organic materials with similar electronic properties to the reference material p-terphenyl. Under strong potassium doping, p-terphenyl was recently reported to exhibit a superconducting transition temperature comparable to those of known high-$T_\mathrm{c}$ cuprates. Resulting from the data-mining exercise, we propose 15 materials as candidates to search for high-$T_\mathrm{c}$} superconductivity (Table~\ref{Tab:results}) which have strong similarities within their DOS to p-terphenyl. Additionally, the presented materials also tend to show similarities within their crystal and molecular structures. Based on the similarities found, we expect similar superconducting mechanisms to take place and suggest experimental verification of the presented results. 

\change{The experimental verification of the candidates presented within this paper under similar conditions as reported for potassium doped p-terphenyl might lead to a better understanding of the prospective high-temperature superconductivity within this material. To date, the mechanism driving the remarkable transition temperature remains under debate. Furthermore, the  crystal structure and chemical composition of the final compound after p-terphenyl was exposed high temperatures as well as highly reactive potassium for several hours remains to be investigated. Therefore, we believe that the field will benefit from investigating additional materials with similar properties.}

\section*{Acknowledgments}
\change{The authors acknowledge helpful discussions with Antonio Bermejo-G\'omez, Philip Hofmann, Bart Olsthoorn and M. Berk Gedik.} The work at Los Alamos is supported by the U.S. Department of Energy, BES E3B7. Furthermore, we are grateful for support from the Swedish Research Council Grant No.~638-2013-9243, the Knut and Alice Wallenberg Foundation, the European Research Council under the European Union’s Seventh Framework Program (FP/2207-2013)/ERC Grant Agreement No.~DM-321031, as well as the Villum foundation through the Villum Center of Excellence for Dirac Materials. The authors acknowledge computational resources from the Swedish National Infrastructure for Computing (SNIC) at the National Supercomputer Centre at Link\"oping University as well as the High Performance Computing Center North.

\bibliography{references.bib}

\end{document}